\documentclass [11pt,a4paper] {article}
\usepackage{amsmath}
\usepackage{amsfonts}
\newcommand{\ed}{\end{document}}
\begin{document}

\title{General solution of classical master equation for reducible gauge theories} 
\author{A.V.Bratchikov\\
 Kuban State
Technological University,\\ Department of Mathematics, \\ Krasnodar, 350072,
Russia}
\date {November, 2011} 
\maketitle

\begin {abstract} We give the general solution to the classical master equation (S,S)=0 for reducible gauge theories. To this aim, we construct a new coordinate system in the extended configuration space 
and transform the equation by changing variables.
Then it can be solved by an iterative method.
\end {abstract}


\section{Introduction}

The classical master equation \cite {Z,BV1,BV2} arises in the Lagrangian approach to gauge theories within the BRST formalism \cite {BRS,T}. It reads 
\begin{eqnarray}
\label{1} (S,S)=0,
\end{eqnarray}
where $S$ is an extended action, and $(.\,,.)$ is an antibracket (for a review of reducible gauge theories see refs. \cite {HT,GPS} and references therein).
The action $S$ also satisfies certain boundary conditions. 
The general solution to this equation for irreducible gauge theories was described in  \cite{VT,BV3}. A similar equation arises in the Hamiltonian approach to the classical BRST charge. 
In both the Lagrangian and Hamiltonian formalisms an important role is played  by the Koszul-Tate differential operator $\delta.$ 
The existence theorem for the BRST charge \cite{FHST}  is based on nilpotency and acyclicity of $\delta.$  By using the results of 
\cite {FHST}, an existence proof of solutions to (\ref{1}) for arbitrary gauge theories was given in \cite{FH}. The authors of ref. \cite{BB} reviewed existence and uniqueness proofs for the extended action in the reducible case and obtained new ones. 

In this paper we present the general solution to the master equation for reducible gauge theories. Our construction is based on a special representation of $\delta$. We find the coordinates in the space of fields and antifields which bring $\delta$ to a standard form.
With respect to the new variables the master equation is simplified and can be solved by an iterative method. 

A similar coordinate system was used for reducing the Koszul-Tate operator and computing cohomology groups of the BRST differential operator in irreducible gauge theories of the Yang-Mills type \cite{BBH}.

The paper is organized as follows.
In section 2, we introduce notations and derive an auxiliary equation. In section 3, we construct new coordinates and transform the operator $\delta$ to a standard form. The 
general solution to the master equation is given in section 4. 
Some examples of reducible theories are discussed in section 5.

In what follows the Grassmann parity and ghost number 
of a function $X$ are denoted by $\epsilon (X)$ and $\mbox{gh}(X),$ 
respectively.  

\section {The classical master equation}

Let $S_0(\phi^{a_0})$ be an action depending on $m_0$ fields $\phi^{a_0} $ of Grassmann parity $\epsilon_{a_0}.$ 
The action is assumed to be gauge invariant 
\begin {eqnarray}\label {U}
S_{0_,a_0}(\phi^{b_0})R^{a_0}_{a_1}(\phi^{b_0})=0,\qquad a_0,b_0=1,\ldots,m_0,\qquad a_1=1,\ldots,m_1.
\end{eqnarray}
The set of generators $R^{a_0}_{a_1}$ forms an algebra, 
$$ R^{a_0}_{b_1,\,b_0}R^{b_0}_{a_1} - (-)^{\epsilon_{a_1}\epsilon_{b_1}}R^{a_0}_{a_1,\,b_0}R^{b_0}_{b_1}
=-R^{a_0}_{c_1}F^{c_1}_{a_1b_1}-S_{0,b_0}E_{a_1b_1}^{a_0b_0}.
$$ 
Here $\epsilon_{a_1}$ is the Grassmann parity of the 
gauge parameter associated with the index ${a_1}$. 

We shall consider a reducible theory 
of $L$-th order. That is, there exist functions 
$$
{R}^{a_k}_{a_{k+1}}(\phi^{b_0}),\qquad  k=0,\ldots,L,\qquad a_k=1,\ldots,m_k, 
$$ 
such that at each stage the $R$'s form a complete set,
$$
R^{a_k}_{a_{k+1}}\lambda^{a_{k+1}}\approx 0 \Leftrightarrow \lambda^{a_{k+1}}\approx R^{a_{k+1}}_{a_{k+2}}\nu^{a_{k+2}},\qquad k=0,\ldots,L-1,
$$
$$
R^{a_L}_{a_{L+1}}\lambda^{a_{L+1}}\approx 0 \Leftrightarrow \lambda^{a_{L+1}}\approx 0,
$$
\begin{eqnarray} \label{r}
R^{a_k}_{ a_{k+1}}R^{a_{k+1}}_{a_{k+2}}
=V^{a_k a_0}_{ a_{k+2}}S_{0,a_0},\qquad  k=0,\ldots,L-1.
\end{eqnarray}
Here $V^{a_0 b_0}_{ a_2}=-(-1)^{\epsilon_{a_0}\epsilon_{b_0}} V^{b_0 a_0}_{ a_2}.$
The weak equality $\approx$ means equality on the stationary surface $${\Sigma \colon \,\,S_{0,a_0}(\phi^{b_0})=0.}$$

The configuration space of the theory is extended by adding the ghost fields 
${C=\left(\phi^{a_1},\phi^{a_2},\ldots, \phi^{a_{L+1}}\right)},$ and the antifields $\phi^*=
(\phi^*_{a_0},\phi^*_{a_1},\ldots, \phi^*_{a_{L+1}}
),$
$$\mbox{gh}(\phi^{a_{k}})=k,\qquad \mbox{gh}(\phi^*_{a_{k}})=-\mbox{gh}(\phi^{a_{k}})-1,\qquad 
\epsilon(\phi^*_{a_k})= \epsilon(\phi^{a_k})+1.$$
We shall seek the extended action $S=S(\phi,\phi^*),$ $\phi=(\phi^{a_0},C),$
in the form of expansions in power series of the ghost fields. 
The antibracket is defined by
$$ 
(X,Y)=\sum_{k=0}^{L+1} \left(
\frac {\delta X} {\delta \phi^{a_k}} \frac {\delta Y} {\delta \phi^*_{a_k}}-(-)^{(\epsilon(X)+1)(\epsilon(Y)+1)}(X\leftrightarrow Y)\right).
$$
Derivatives with respect to antifields are always understood as left, while with respect to fields as right.   

Eq. \eqref {1} is supplied by
the following conditions
\begin{eqnarray} \label{axu}
\epsilon (S)=0, \qquad \mbox {gh} (S)=0,
\end{eqnarray}
\begin{eqnarray*} 
\left. S \right|_{C=\phi^* 
=0}=S_{0},\qquad \left.\frac {\delta^2 S} {\delta \phi^*_{a_{k-1}}\delta \phi^{a_k}}\right|_{C=\phi^*
=0} =R_{a_k}^{a_{k-1}},\qquad k=1,\ldots,L+1.  
\end{eqnarray*}
We assume that $S$ is a local functional.

One can write
\begin{eqnarray} \label{us2}
S=S_0+S_1+K,\qquad K = \sum_{n\geq 2} S_n,\qquad  
 S_n \sim  C^{n},
\end{eqnarray}
\begin{eqnarray} \label {sumo}
S_1= \sum_{k=1}^{L+1}\left(\phi^*_{a_{k-1}}R^{a_{k-1}}_{a_{k}}+M_{a_{k}}\right)\phi^{a_{k}},
\end{eqnarray}
where $M_{a_{k}}=M_{a_{k}}(\phi^{a_0},\phi^*_{a_1},\ldots,\phi^*_{a_{k-2}}).$  
Eq. \eqref{axu} implies that 
\begin{eqnarray*} 
\left.M_{a_{k}}\right|_{\phi^*=0}=0, \qquad \left. K \right|_{\phi^*=0}=0.
\end{eqnarray*}

Let ${\cal V}$ be the space of the local functionals depending on $(\phi,\phi^{*})$ 
which vanish on $\Sigma$ at $\phi^{*}=0.$ It is 
easily verified that 
$S_n,n\geq 1,$ as well as $(S,S),$ belong to ${\cal V}.$
 
Substituting \eqref {us2} in \eqref {1}
one obtains 
\begin{eqnarray} \label{ru}
\delta S_{1}=0,
\end{eqnarray}
\begin{eqnarray} \label{fund}
\delta K = D,
\end{eqnarray}
where
\begin{eqnarray} \label{de}
\delta = S_{0,a_0}
\frac{\delta }{\delta \phi^*_{a_0}}+
\sum_{k=1}^{L+1} \left(\phi^*_{a_{k-1}} R^{a_{k-1}}_{a_k} + M_{a_{k}}\right)\frac{\delta }{\delta \phi^*_{a_k}},
\end{eqnarray}
 \begin{eqnarray*}
 \label{r_v2}-D=B+AK+\frac 1 2 (K,K),\qquad B=S_{1,a_0} R^{a_0}_{a_1}\phi^{a_1},
\end{eqnarray*}
and the operator $A$ is defined by  
\begin{eqnarray*} 
AX=S_{1,a_0} \frac{\delta X }{\delta \phi^*_{a_0}}+(-1)^{\epsilon(X)}\sum_{k=0}^{L+1}
\frac{\delta X}{\delta \phi^{a_k}}\frac{\delta {S_1} }{\delta \phi^*_{a_k}}.
\end{eqnarray*} 
Eq. \eqref{ru} is equivalent to   
\begin{eqnarray} \label{r22}
\delta^2=0.
\end{eqnarray}

Assume that (\ref{ru}) holds. Then $(S,S)=G,$ where $G$ is the difference between the left-hand and right-hand sides of (\ref {fund}), 
\begin {eqnarray*}
G=\delta K+B+AK+\frac 1 2 (K,K).
\end {eqnarray*}
From the Jacobi identity $(S,(S,S))=0$ it follows that $(S,G)=0,$ or equivalently
\begin {eqnarray}
\label{omf}
\delta G+ A G + (K,G)=0. 
\end {eqnarray}

\section {Reduction of 
$\delta$}

For $k=L$ eq. (\ref{r}) reads 
\begin{eqnarray} \label{r1}
R_{a'_{L}}^{a_{L-1}}
R_{a_{L+1}^{\phantom {\prime}}}^{a'_{L}}+
R^{a_{L-1}}_{a''_{L}}R^{a''_{L}}_
{a_{L+1}^{\phantom {\prime}}
}\approx 0,
\end{eqnarray}
where
$a'_{L},a''_{L}$ are index sets, such that $a'_{L}\cup\, a''_{L}= a_{L},$ $|a'_{L}|=|a_{L+1}|,$ 
and ${{\rm rank}\,\left.R^{a'_{L}}_{a_{L+1}^{\phantom {\prime}}}\right|_\Sigma =|a_{L+1}|.}$~{\footnote{For an index set $i= \{i_1,i_2,\ldots,i_n\}, |i|=n.$}} It follows from  (\ref{r1}) 
that $${{\rm rank}\, \left.R^{a_{L-1}}_{a_L^{\phantom {\prime}}}\right|_\Sigma=
{\rm rank}\,\left. R^{a_{L-1}}_{a''_{L}}\right|_\Sigma=| a''_{L}|.}$$

One can split the index set $a_{L-1}$ as
$a_{L-1}=a'_{L-1}\cup\, a''_{L-1},$ such that ${|a'_{L-1}|=|a''_{L}|}$ and ${\rm rank}\, \left. R^{a'_{L-1}}_{a''_{L}}\right|_\Sigma=|a''_{L}|.$
For $k=L-1$ eq. (\ref{r}) implies 
\begin{eqnarray*} 
R^{a_{L-2}^{\phantom {\prime}}}_{a'_{L-1}}
R_{a''_{L}}^{a'_{L-1}}+
R^{a_{L-2}^{\phantom {\prime}}
}_{a''_{L-1}}R^{a''_{L-1}}_{a''_{L}}\approx 0.
\end{eqnarray*}
From this it follows that $${\rm rank}\, \left. R^{a_{L-2}}_{a_{L-1}^{\phantom {\prime}}}\right|_\Sigma = {\rm rank}\, \left.R^{a_{L-2}}_{a''_{L-1}}\right|_\Sigma=| a''_{L-1}|.$$

Using induction on $k,$ one can obtain a set of matrices $R^{a_{k-1}}_{a''_{k}},$ ${k=1,\ldots,L+1,}$
satisfying $${\rm rank}\, 
\left.
R^{a_{k-1}}_{a''_{k}}
\right
|_\Sigma={\rm rank}\, \left.R^{a_{k-1}}_{a_k^{\phantom {\prime}}}\right|_\Sigma=|a''_k|,$$
and a set of nonsingular matrices $R^{a'_{k-1}}_{a''_{k}},k=1,\ldots,L+1,$ where
$a'_{k}\cup\, a''_{k}=a_{k}.$

Eq. \eqref{U} implies 
\begin{eqnarray*}
S_{0,b_0^{\phantom {\prime}}a'_0}R^{a'_0}_{a''_1} + S_{0,b_0^{\phantom {\prime}}a''_0}R^{a''_0}_{a''_1} \approx 0,
\end{eqnarray*}
and therefore $${{\rm rank}\,\left. S_{0,b_0^{\phantom {\prime}}a_0^{\phantom {\prime}}}\right|_\Sigma={\rm rank}\,\left. S_{0, b_0^{\phantom {\prime}}a''_0}\right|_\Sigma=|a''_{0}|.}$$

For $a''_{k+1}\subset a_{k+1},$ $k=0,\ldots,L-1,
$ we define an embedding ${f(a''_{k+1})=a''_{k+1}\subset a_{k},}$ ${f(a''_{L+1})=a_{L+1}\subset a_{L}.}$ Thus, for example, $(\phi^*_{f(a''_{k+1})})\subset (\phi^*_{a_k}).$  
Let $\alpha_{k},$ ${k=0,\ldots,L,}$ be defined by ${a_{k}= f(a''_{k+1})\cup \alpha_{k}.}$ One can write $\alpha_{k}=g(a''_{k})$ for some function $g,$ since 
$|a''_{k}|=|\alpha_{k}|.$ 

{\it Lemma.} The nilpotent operator $\delta$ is reducible to the form 
\begin{eqnarray*} 
\delta= \phi'_{a''_0} \frac{\delta}{\delta \phi^{*\prime}_{g(a''_0)}}
+\sum_{k=1}^{L+1} \phi^{*\prime}_{f(a''_k)}\frac{\delta}{\delta \phi^{*\prime}_{g(a''_k)}},
\end{eqnarray*}
by the change of variables:
$\left(\phi^{a_0},\phi^*\right)\to
\left(\phi'_{a_0},\phi^{*\prime}\right),$
\begin{eqnarray*}
\phi'_{a''_0} =  S_{0,a''_0}, \qquad \phi'_{a'_0} = \phi^{a'_0},
\end{eqnarray*}
\begin{eqnarray}
\label{cu}
\phi^{*\prime}_{f(a''_{k+1})}=\phi^*_{a_k}
R^{a_{k}}_{a''_{k+1}}+ 
M_{a''_{k+1}},\qquad 
\phi^{*\prime}_{\alpha_{k}}={\phi^*}_{g^{(-1)}(\alpha_{k})},
\end{eqnarray}
\begin{eqnarray*}
\phi^{*\prime}_{a_{L+1}} = {\phi^*}_{a_{L+1}},
\end{eqnarray*}
where $k=0,\ldots,L,$ $g(a''_{L+1})=a_{L+1}.$

To prove this statement we first observe that the matrices ${(\tilde S_{a_0 b_0})= (S_{0,a''_0b''_0},\delta _{a'_0b'_0}),}$ $(\tilde R^{b_k}_{a_k})= (R^{b_{k}}_{f(a''_{k+1})},\delta ^{b_{k}}_{\alpha_{k}}),$ $k=0,\ldots, L,$ are invertible, and therefore transformation \eqref{cu} is nonsingular.
For $0~\leq~s~\leq~L$   
$$ \label{de10}
\frac{\delta }
{\delta \phi^*_{a'_s}}= 
\sum_{k=0}^L\frac{\delta (\delta \phi^*_{a''_{k+1}})}
{\delta \phi^*_{a'_s}}\frac{\delta }{\delta \phi^{*\prime}_{f(a''_{k+1})}},\qquad
\frac{\delta}
{\delta \phi^*_{a''_s}}= \sum_{k=0}^L\frac{\delta
(\delta \phi^*_{a''_{k+1}})}{\delta \phi^*_{a''_s}}\frac {\delta}{\delta \phi^{*\prime}_{f(a''_{k+1})}}+
\frac{\delta}{\delta \phi^{*\prime}_{g(a''_s)}}.
$$
Substituting this in \eqref {de}, we get 
$$
\delta = \sum_{k=0}^{L}\left(\delta^2 \phi^*_
{a''_{k+1}}\frac {\delta } {\delta \phi^{*\prime}_{f(a''_{k+1})}}+ 
\delta \phi^*_{a''_k}\frac{\delta }{\delta \phi^{*\prime}_{g(a''_k)}}\right)+\delta \phi^*_{a_{L+1}}\frac{\delta }{\delta \phi^{*\prime}_{a_{L+1}}}.$$
The result then follows from \eqref{r22} and \eqref{cu}.

Eqs. \eqref {cu} are  solvable with respect to the original variables 
and can be represented as  
\begin{eqnarray*}\label{cucu}
\phi^{a_0}= \phi^{a_0}(\phi'_{b_0}),\qquad \phi^*_{a_k}=\phi^{*}_{a_k}(\phi'_{a_0},
\phi^{*\prime}_{a_0},\ldots,\phi^{*\prime}_{a_k}),\qquad  k=0,\ldots,L+1.
\end{eqnarray*}
Here we have used the fact that the $\phi^*_{a_k}$ depends 
only on the functions $\phi^{*\prime}_{a_s}$ with $s\leq k.$
Assume that the functions $\phi^{a_0}(\phi'_{b_0})$ have been constructed.
Then from (\ref{cu})  
\begin{eqnarray*}\label{cucu22}
\phi^{*}_{a'_k}=
(\phi^{*\prime}_{f(a''_{k+1})}- 
\phi^{*\prime}_{g(a''_k)} R^{\prime a''_{k}}_{a''_{k+1}}- M^{\prime\phantom{a_k}}_{a''_{k+1}})
(R^{\prime(-1)})^{a''_{k+1}}_{a'_k},
\qquad \phi^{*}_{a''_k}=\phi^{*\prime}_{g(a''_k)},
\end{eqnarray*}
\begin{eqnarray}\label{cucu22} 
\phi^*_{a_{L+1}}= \phi^{\prime*}_{a_{L+1}},
\end{eqnarray}
where $k=0,\ldots,L,$
\begin{eqnarray*}
R^{\prime a_k}_{a_{k+1}}(\phi'_{a_0})=R^{a_k}_{a_{k+1}}(\phi^{a_0}),\qquad
M'_{a_k}(\phi'_{a_0},\phi^{*\prime}_{a_0},\ldots,\phi^{*\prime}_{a_{k-2}})=
M^{\phantom{b_k}}_{a_k}(\phi^{a_0},\phi^*_{a_0},\ldots,\phi^*_{a_{k-2}}).
\end{eqnarray*}

With respect to the new coordinate system the condition $X\in {\cal V}$ 
implies 
$$\left. X \right|_{\phi'_{a''_0}=\phi^{*\prime}=0}=0.$$
For $X'_i(\phi'_{a_0},\phi^{*\prime},C),$ $i=1,2,$
we define 
$$(X'_1,X'_2)'=(X_1,X_2),$$
where 
$$X_i(\phi^{a_0},\phi^*,C)=X'_i(\phi'_{a_0},\phi^{*\prime},C).$$
Notice that the variables ${\phi'_{a''_0},\phi^{*\prime}_{g(a''_0)},\phi^{*\prime}_{f(a''_1)},\phi^{*\prime}_{g(a''_1)},\ldots,
\phi^{*\prime}_{f(a''_{L+1})},\phi^{*\prime}_{g(a''_{L+1})}}$ are independent.

\section {The extended action}
\paragraph {Solution of the equation $\mathbf {\delta^2=0}$.}

Eq. (\ref{r22}) is equivalent to the recurrent relations ${M_{a_1}=0,}$
\begin{eqnarray} \label {f}
\delta M_{a_k}=-(\phi^{*}_
{a_{k-2}}R_{a_{k-1}}^{a_{k-2}}+M_{a_{k-1}})R_{a_k}^{a_{k-1}},\qquad k=2,\ldots,L+1.  
\end{eqnarray}
One can replace $\delta$ by $\delta_{k},$
\begin{eqnarray*} \label{f0}
\delta_k = S_{0,\,a_0}
\frac{\delta }{\delta \phi^*_{a_0}}+
\sum_{s=1}^{k-2} \left(\phi^*_{a_{s-1}} R^{a_{s-1}}_{a_s} + M_{a_{s}}\right)\frac{\delta }{\delta \phi^*_{a_s}},
\end{eqnarray*}
since $M_{a_k}$ does not depend on $\phi^{*}_
{a_{s}},s>k-2.$
The operator $\delta_{k}$ and right-hand side of (\ref {f})
only involves the functions $M_{a_s}$ with $s<k.$ 

Assume that the functions $M_{a_s},s<k,$ have been constructed.
Changing variables in \eqref {f} $
(\phi^{a_0},\phi^*_{a_0},\ldots, \phi^*_{a_{k-2}})\to
(\phi'_{a_0},\phi^{*\prime}_{a_0},\ldots, \phi^{*\prime}_{a_{k-2}}),
$
we get 
\begin{eqnarray} \label{f1}
\delta_{k}  M'_{a_k}= D'_{a_k},
\end{eqnarray}
where
$$\delta_{k}= \phi'_{a''_0} \frac{\delta }{\delta \phi^{*\prime}_{g(a''_0)}}
+\sum_{s=1}^{k-2} \phi^{*\prime}_{f(a''_s)}\frac{\delta }{\delta \phi^{*\prime}_{g(a''_s)}}, \qquad
D'_{a_k}= - (\phi^{*}_
{a_{k-2}}R_{a_{k-1}}^{\prime a_{k-2}}+ M'_{a_{k-1}})R_{a_k}^{\prime a_{k-1}},
$$
$$ 
\phi^{*}_
{a_{k-2}}= \phi^{*}_{a_{k-2}}(\phi'_{a_0},\phi^{*\prime}_{a_0},\ldots,\phi^{*\prime}_{a_{k-2}}).
$$

Let $n_{k}$ 
be the counting operator 
$$
n_{k} = \phi'_{a''_{0}}\frac{\delta }{\delta \phi'_{a''_{0}}}+ \phi^{*\prime}_{g(a''_{0})}\frac{\delta}{\delta 
\phi^{*\prime}_{g(a''_{0})}}+ \sum_{s=1}^{k-2}\left(\phi^{*\prime}_{f(a''_{s})}\frac{\delta}{\delta \phi^{*\prime}_{f(a''_{s})}}
+  \phi^{*\prime}_{g(a''_{s})}\frac {\delta} {\delta \phi^{*\prime}_{g(a''_{s})}}\right),
$$
and let 
$$ 
\sigma_{k}= \phi^{*\prime}_{g(a''_0)} \frac{\delta}{\delta \phi'_{a''_0}}+
\sum_{s=1}^{k-2} \phi^{*\prime}_{g(a''_{s})}\frac{\delta }{\delta \phi^{*\prime}_{f(a''_{s})}}
.
$$
One can directly verify that 
\begin{eqnarray} \label{us4} 
\delta^2_k=\sigma^2_{k}=0,\qquad \delta_{k}\sigma_{k}+\sigma_{k} \delta_{k}=n_{k},\qquad n_{k}\delta_{k}=\delta_{k} n_{k} , 
\qquad n_{k}\sigma_{k}=\sigma_{k} n_{k}.
\end{eqnarray}

Let ${\cal V}_{k},$ $2\le k \le L+3,$ be the subspace of ${\cal V}$ which consists of
the functionals depending only on $(\phi'_{a_0},\phi^{*\prime}_{a_0}, \ldots, \phi^{*\prime}_{a_{k-2}}).$ 
Notice that ${\cal V}_{L+3}={\cal V}.$ 
The space ${\cal V}_k$ splits as 
$$ 
{\cal V}_k= {\cal V}_{k}^{(0)}\oplus\widetilde{\cal V}_k, \qquad 
\widetilde{\cal V}_k= {\cal V}_{k}^{(1)} \oplus{\cal V}_{k}^{(2)} \oplus \ldots,
$$
with  $
n_k X=nX$ for $X\in {\cal V}_{k}^{(n)}.$ 
It is clear that  
\begin {eqnarray}
\label{xoss}
{\cal V}_{k}^{(0)}=\{ \Phi\in {\cal V}_k\,|\, \Phi=\Phi( \phi'_{a'_0},
\phi^{*\prime}_{f(a''_{k-1})})
\},\quad k\ne L+3,\qquad {\cal V}_{L+3}^{(0)}=0.
\end {eqnarray}

We define $n_k^+: {\cal V}_k\to{\cal V}_k$ by 
$$
n_k^+X =
\left \{
\begin {array}{rcl} 
n_k^{(-1)}X,&\phantom {h} & X \in \widetilde {\cal V}_k;\\
0,\phantom {X}&\phantom {k} &  X \in {\cal V}_k^{(0)}.\\
\end{array}
\right.
$$
where $n_k^{(-1)}: \widetilde{\cal V}_k\to\widetilde{\cal V}_k$ is given by
$$
n_k^{(-1)}X =\frac 1 {n}X,\qquad
X \in {\cal V}_k^{(n)},\qquad  n>0.$$
Then  $\delta_{k}^{{+}}=\sigma_{k} n_{k}^{+}$ is a generalized inverse of $\delta_{k}$: 
\begin{eqnarray} \label{mk} 
\delta_{k}\delta_{k}^{+}\delta_{k}=\delta_{k},
\qquad \delta_{k}^{+}\delta_{k}\delta_{k}^{+}=\delta_{k}^{+}
.
\end{eqnarray} 
We shall denote $\delta^{+}=\delta_{L+3}^{+}.$

Eq. (\ref{r}) takes the form 
\begin{eqnarray*} 
R^{\prime a_k}_{ a_{k+1}}R^{\prime a_{k+1}}_{a_{k+2}}
=V^{\prime a_k a'_0}_{ a_{k+2}}S'_{a'_0}+ V^{\prime a_k a''_0}_{ a_{k+2}}\phi'_{a''_0},
\qquad  k=0,\ldots,L-1,
\end{eqnarray*}
where 
$$V^{\prime a_0b_0}_{a_2}(\phi'_{a_0})=V^{a_0b_0}_{a_2}(\phi^{a_0}),\qquad
S'_{b'_0}(\phi'_{a_0})=S_{0,\,b'_0}(\phi^{a_0}).$$ 
It follows from (\ref{U}) that 
$$S'_{a'_0}=- \phi'_{a''_0}U^{\prime a''_0}_{a'_0}, \qquad U^{\prime a''_0}_{a'_0}=R^{\prime a''_0}_{a''_1}(R^{\prime(-1)})^{a''_1}_{a'_0},$$
and hence 
\begin{eqnarray} \label{au}R^{\prime a_k}_{ a_{k+1}}R^{\prime a_{k+1}}_{a_{k+2}}\in \widetilde{\cal V}_2.
\end{eqnarray}
We assume that $M'_{a_s}\in \widetilde {\cal V}_s,$ $s<k.$ Then, by (\ref {au}), 
$D'_{a_k}\in \widetilde{\cal V}_k.$
One can directly verify that $\delta_k  D'_{a_k}=0,$ or equivalently
$\delta_{k}\delta_{k}^{+}D'_{a_k}=D'_{a_k}.$
Then the general solution to (\ref {f1}) for $2\le k \le L$
is given by
\begin{eqnarray} \label{o10}
M'_{a_k} = Y'_{a_k} + \delta_{k}^{{+}}D'_{a_k},
\end{eqnarray} 
where $Y'_{a_k}\in \widetilde {\cal V}_k $ is an arbitrary cocycle, $\delta_k Y'_{a_k}=0,$ subject only to the restrictions 
\begin{eqnarray} \label{o23}\epsilon(Y'_{a_k})=\epsilon(M'_{a_k}),\qquad \mbox{\rm gh}(Y'_{a_k})=\mbox{\rm gh}(M'_{a_k}).
\end{eqnarray}
By its construction, $M'_{a_k}\in \widetilde {\cal V}_k.$
The function $M'_{a_{L+1}}$ is given by (\ref {o10}), (\ref {o23}),
where the cocycle $Y'_{a_{L+1}}$ belongs to ${\cal V}_{L+1}.$  

\paragraph {Higher orders.}
Consider eq. \eqref{fund}. Changing variables $
{(\phi^{a_0},\phi^{*}) \to
(\phi'_{a_0},\phi^{*\prime}),}$ 
we get
\begin {eqnarray} \label{to}
\delta K^{\prime} = D^{\prime},
\end{eqnarray}
where 
\begin {eqnarray*}
-D'= B'
+ A K'+\frac 1 2 (K',K')'
\end {eqnarray*}
$$B^{\prime}(\phi',\phi^{*\prime},C)= B(\phi, \phi^{*}),\qquad
K^{\prime}(\phi',\phi^{*\prime},C)= K(\phi, \phi^{*}).
$$

Applying $\delta\delta^+$ to eq. (\ref{to}) and using (\ref{mk}) we have 
$$\delta K'=\delta\delta^{+}D',$$ from which it follows that   
\begin {eqnarray} \label{tor}
K^{\prime}= Y' + \delta^{+}D^{\prime},
\end {eqnarray}  
\begin {eqnarray}
\label{rssr}
Y'\in {\cal V}, \qquad
\delta Y'=0,\qquad \epsilon(Y')= 0,\qquad \mbox{\rm gh} (Y)= 0.
\end {eqnarray}

Let $\langle .\,,. \rangle: {\cal V}^2\to {\cal V}  $ be defined by
\begin{eqnarray*} \label {or}
\langle X'_1,X'_2 \rangle = -\frac 1 2 (I+\delta^+A)^{-1}\delta^{+}\left(( X'_1,X'_2 )'+( X'_2,X'_1)'\right),
\end{eqnarray*} 
where $I$ is the identity map, and
$$(I+\delta^+A)^{-1}=\sum_{m\geq 0}(-1)^m(\delta^+A)^m.$$
One can rewrite  (\ref{tor}) as 
\begin {eqnarray}
\label{omee}
K' = K'_0 + \frac 1 2 \langle K',K' \rangle,
\end {eqnarray}
where 
\begin {eqnarray*}
\label{mo}
K'_0 = (I+\delta^+A)^{-1}\left(Y'- \delta^{+}B'\right),
\end {eqnarray*}
Eq. (\ref {omee}) can be iteratively solved as:
\begin {eqnarray}
\label{mex}
K' = K'_0 + \frac 1 2 \langle K'_0,K'_0 \rangle +\ldots 
\end {eqnarray}

To prove that the solution to eq. (\ref {tor}) satisfies (\ref {to}) we shall use the approach of ref. \cite {F}. 
From (\ref{us4}) and the second relation of (\ref{xoss}) it follows that for any $X
\in {\cal V},$  
\begin {eqnarray}
\label{osas}
X=
\delta^+
\delta 
X+
\delta
\delta^+
X.
\end {eqnarray}
Changing variables in (\ref{omf}) from $
(\phi^{a_0},\phi^*)$ to
$(\phi'_{a_0},\phi^{*\prime}),$ we get
\begin {eqnarray}
\label{oss}
\delta G'+ A G' + (K',G')'=0, 
\end {eqnarray}
where 
\begin {eqnarray}
\label{omt}
G'=\delta K'+B'+AK'+\frac 1 2 (K',K')'.
\end {eqnarray}
Consider eq. (\ref{oss})
and the condition
\begin {eqnarray}
\label{ossa}
\delta^+G'=0, 
\end {eqnarray}
where $K'$ is the solution to (\ref{tor}). 
Applying $\delta^+$ to eq. (\ref{oss}) and using (\ref{ossa}) we get 
\begin {eqnarray}
\label{ossr}
 G'=-\delta^+ (A G' + (K',G')'), 
\end {eqnarray}
since $G'\in {\cal V}.$ 
From (\ref{ossr}) it follows that $G'=0.$

The solution to (\ref{tor}) satisfies the condition 
\begin {eqnarray*}
\delta^+K' = \delta^+Y', 
\end {eqnarray*}
since $(\delta^+)^2=0.$
By using 
(\ref{osas}), we have
 \begin {eqnarray*}
 K'=\delta^+\delta K' + \delta\delta^+Y'. 
\end {eqnarray*}
From this, (\ref{osas}) and (\ref{rssr}) it follows that
\begin {eqnarray}
\label{osm}
 K'=\delta^+\delta K' + Y'. 
\end {eqnarray}
To check (\ref{ossa}), we have 
\begin {eqnarray*}
\label{omt}
\delta^+G'=\delta^+\delta K'+\delta^+(B'+AK'+\frac 1 2 (K',K')'),
\end {eqnarray*}
and therefore by 
(\ref{osm}) and (\ref{tor}), $\delta^+G'=0.$  

Changing variables in (\ref {o10}) and (\ref {mex}) 
from $(\phi'_{a_0},\phi^{*\prime})$ to
$(\phi^{a_0},\phi^{*}),$ one can obtain $S_1(\phi,\phi^*)$ and $K(\phi,\phi^*).$ 
The construction of the inverse transformation may be problematic in a field theory.
It should be noted, however, that the extended action can be computed by using formal expressions \eqref {cucu22}, because the result does not depend on the choice of an auxiliary coordinate system.

\section {Reducible abelian gauge theories}

In this section we discuss a class of reducible bosonic theories including, in particular, 
the antisymmetric tensor and abelian $p$-form gauge theories.

It follows from (\ref{omee}) that if $\delta^{+}B'=Y'$ then $K=0,$ and therefore 
\begin{eqnarray} 
\label{us22}
S=S_0+S_1.
\end{eqnarray}
Let us consider the case of $B'=0,$ $Y'=0.$ The equation $B=0$ is equivalent to the relations
\begin{eqnarray} \label{us23}
R^{a_0}_{b_1,b_0}R^{b_0}_{a_1} - R^{a_0}_{a_1,b_0}R^{b_0}_{b_1}=0,
\end{eqnarray}
\begin{eqnarray} \label{us25}
R^{a_{k-1}}_{a_k,b_0}R^{b_0}_{a_1}=0,\qquad k=2,\ldots,L+1, 
\end{eqnarray}
\begin{eqnarray} \label{us24}
M_{a_k,a_0}
R^{a_0}_{a_1}=0.
\end{eqnarray}
Eq. (\ref{us23}) tells us that the gauge algebra is abelian.

Let us consider the case of reducible abelian gauge theories of the first order.
The reducibility identities read   
\begin{eqnarray} \label{rx}
R^{a_0}_{ a_{1}}R^{a_{1}}_{a_{2}}=V^{a_0b_0}_{a_2}S_{0_,\,b_0},\qquad V^{a_0b_0}_{a_2}=-V^{b_0a_0}_{a_2}.
\end{eqnarray}
We assume that $V^{a_0b_0}_{a_2}$ are constants, and (\ref{us25}) is satisfied.

It is easily verified that for $k=2$ the general solution to 
(\ref{f}) is given by 
\begin{eqnarray} \label{ramg}
M_{a_2}= Y_{a_2}+
 \frac 1 2 \phi^{*}_{a_0}V^{a_0b_0}_{a_2}\phi^{*}_{b_0}, 
\end{eqnarray}
where  $Y_{a_2}= Y_{a_2}(\phi^{a_0},\phi^{*}_{a_0})$ is a cocycle. 
If $Y_{a_2,a_0}
R^{a_0}_{a_1}=0,$ then $M_{a_2}$ 
satisfies (\ref{us24}). We can set $Y_{a_2}=0.$
Then
$$
S=S_0+\phi^*_{a_0}R^{a_0}_{a_1}\phi^{a_1}+
(\phi^*_{a_{1}}R^{a_{1}}_{a_{2}}+ \frac 1 2 \phi^{*}_{a_0}V^{a_0b_0}_{a_2}\phi^{*}_{b_0})\phi^{a_{2}}.
$$ 

An example of such a theory is the antisymmetric tensor gauge theory \cite{FT} whose dynamics is described by the action 
\begin{eqnarray} \label{rams}S_0=\int d^4x\left (\frac 1 2 A^a_\mu A^{a\mu}- \frac 1 2  B^a_{\mu\nu}F^{a\mu\nu}\right),
\end{eqnarray}
where 
$ F^{a}_{\mu\nu}=
\partial_\mu 
A^a_\nu - 
\partial_\nu A^a_\mu
-f^a
_{bc}
A^b_\mu
A^c_\nu,$ $f^a
_{bc}$ are the structure constants of a semi-simple compact Lie algebra. 

The action is invariant under the gauge transformations
\begin{eqnarray} 
\delta B^{a\mu\nu}=R^{a\mu\nu}_{b\lambda}\xi^b_\sigma,\qquad \delta A^a_\mu=0,
\end{eqnarray}
where  
\begin{eqnarray*} 
R^{a\mu\nu}_{b\lambda}=\varepsilon^{\mu\nu\rho\sigma}\eta_{\sigma\lambda}
\nabla^a_{\rho b}, \qquad
\nabla^{a}_{\mu b}=\partial_\mu\delta^a_b-f^a_{cb}A^c_\mu, 
\end{eqnarray*}
$\eta_{\sigma\lambda}=diag (-1,1,1,1).$
It is easily seen that the gauge algebra is abelian. 
The equations of motion are 
\begin{eqnarray} 
\label{ramu}
\frac {\delta S_0}{\delta B^{a\mu\nu}}=-F_{a\mu\nu}=0,\qquad 
\frac {\delta S_0}{\delta A^{a\mu}}=A_{a\mu}+\nabla^{\nu a}_{b} B^{b}_{\nu\mu}=0.
\end{eqnarray}
This theory is on-shell first stage reducible, since  
\begin{eqnarray} 
\label{ramsis}
R^{c\mu\nu}_{b\lambda}R^{b\lambda}_a=\frac 1 2 \varepsilon^{\mu\nu\rho\sigma} f^c_{ab}F^{b}_{\rho\sigma},
\end{eqnarray}
and $R^{b\lambda}_a=\nabla^{\lambda b}_a$ are independent. It is easily seen that (\ref{us25}) holds.

We have 
\begin{eqnarray*} 
\phi^{a_0}=(B^a_{\mu\nu}(x), A^b_\lambda(y)),\qquad 
a_0=(a,\mu,\nu,x)\cup  (b,\lambda,y),\qquad \mu<\nu,
\end{eqnarray*}
\begin{eqnarray*} 
a'_0=(a,i,j,x),\qquad 
a''_0=(a,k,3,x)\cup (b,\lambda,y),\qquad 0\leq i,j,k\leq 2,\qquad i<j,
\end{eqnarray*}
\begin{eqnarray*} 
\phi^{a_1}=c^a_\mu(x),\qquad a_1=(a,\mu,x),\qquad a'_1=(a,3,x),\qquad a''_1=(a,i,x),
\end{eqnarray*} 
\begin{eqnarray*} 
 \phi^{a_2}=c^a(x), \qquad a_2=(a,x),
\end{eqnarray*} 
\begin{eqnarray*}
\phi^*_{a_0}=(B^{*a}_{\mu\nu}(x), A^{*b}_\lambda(y)), \qquad \phi^*_{a_1}=c^{*a}_\mu(x),\qquad \phi^*_{a_2}=c^{*a}(x).
\end{eqnarray*}

Let $\nabla^{(-1)a}_{\phantom{(-1)}b}(x,y)$ denote an inverse of $\nabla^a_{b}\delta(x-y),$ $\nabla^a_{b}=\nabla^a_{3b},$
\begin{eqnarray*} 
\nabla^{a}_{c}(x)\nabla^{(-1)c}_{\phantom{(-1)}b}(x,y)=\delta^a_b\delta(x-y).
\end{eqnarray*}
We impose the boundary conditions that all the fields and antifields vanish at $x_3\to -\infty.$
Then the inverse $\nabla^{(-1)a}_{\phantom{(-1)}b}(x,y)$ is unique. It can be obtained by iterating the equation
\begin{eqnarray*} 
\nabla^{(-1)a}_{\phantom{(-1)}b}
(x,y)=
\delta^a_b(\delta^3\theta)(x-y)- (\delta^3\theta H^a_c
\nabla^{(-1)c}_{\phantom{(-1)}b})(x,y),
\end{eqnarray*}
where 
\begin{eqnarray*}
(\delta^3\theta)(x)=\delta(x_0)\delta(x_1)\delta(x_2)\theta(x_3),\qquad H^a_{b}(x,y)=f^a_{cb}A^c_3(x)\delta(x-y),
\end{eqnarray*}
$\theta(x_3)=1$ if $x_3\geq 0$ and $\theta(x_3)=0$ otherwise.

The operator 
\begin{eqnarray*} 
R^{a'_1}_{a_2}=R^{b3}_{a}=\nabla^{b}_a
\end{eqnarray*}
is invertible. Here and in what follows we omit space-time indexes. It is easily verified that 
\begin{eqnarray*} 
R^{a'_0}_{a''_1}=R^{aij}_{bk}=\varepsilon^{lij}\eta_{lk}\nabla^a_{b},\qquad i<j,
\end{eqnarray*}
is also invertible.

In accordance with (\ref{cu}), (\ref{ramu}) the substitution  $(B^{a}_{\mu\nu},A^b_{\lambda})
\to (B^{\prime a}_{\mu\nu},A^{\prime b}_{\lambda})$ looks like
\begin{eqnarray*} 
B^{\prime a}_{i3}=-F^a_{i3},\qquad A^{\prime a}_\nu=A^a_\nu+\nabla^{\mu a}_{b}
B^{b}_{\mu \nu},
\qquad B^{\prime a}_{ij}=B^{a}_{ij}.
\end{eqnarray*}
Solving these equations with respect to 
$(B^{a}_{\mu\nu},
A^{b}_\lambda),$
we get 
\begin{eqnarray*} 
\label{rama}
B^{a}_{ij}=B^{\prime a}_{ij},\qquad A^{a}_3=A^{\prime a}_3,\qquad
A^{a}_i= \nabla^{\prime (-1)a}_{\phantom{\prime (-1)}b}   (\partial_i A^{\prime b}_3+B^{\prime b}_{i3}), 
\end{eqnarray*}
\begin{eqnarray*} 
B^{a}_{j3}=\nabla^{\prime (-1)a}_{\phantom{\prime (-1)}b}(
A^{b}_{j}-A^{\prime b}_{j}+\nabla^{\prime i b}_c B^{\prime c}_{i j}). 
\end{eqnarray*}
Here 
$$A^{a}_{j}=A^{a}_{j}(A^{\prime b}_3, B^{\prime c}_{i3}),\qquad
\nabla^{\prime a}_{\mu b}(A^{\prime c}_\nu)=
\nabla^{a}_{\mu b}(A^c_\mu).$$ 

Let us now consider the substitution $(B^{*a}_{\mu\nu},A^{*b}_{\lambda})
\to (B^{*\prime a}_{\mu\nu},A^{*\prime b}_{\lambda}):$
\begin{eqnarray} 
\label{ramss}
B^{*\prime}_{f(a,i)}=B^*_{b\mu\nu}
\varepsilon^{\mu\nu\lambda j}\eta_{ji}{\nabla}^b_{\lambda a}
,\qquad B^{*\prime a}_{j3}=
B^{* a}_{j3},\qquad A^{*\prime a}_\mu=A^{*a}_\mu,
\end{eqnarray}
where \begin{eqnarray*} 
f^{(-1)}(a,i,j)=(a,|\varepsilon^{ij0}|+ \sum_{k>0}|\varepsilon^{ijk}k|),\qquad i<j.
\end{eqnarray*}
From (\ref{ramss}) it follows that $$B^{* a}_{j3}=
B^{*\prime a}_{j3},\qquad A^{*a}_\mu=A^{*\prime a}_\mu,$$
$$ 
B^{*}_{aij}=(B^{*\prime}_{ci3}\nabla^c_{jb}-B^{*\prime}_{cj3}\nabla^c_{ib}-\frac 1 2  \varepsilon_{lij} \eta^{lk}  B^{*\prime}_{f(b,k)})
\nabla^{(-1)b}_{\phantom{(-1)}a}. 
$$

Let us derive an expression for $M_{a_2}=M_{a}$ from (\ref{o10}). By using (\ref{ramsis}), we get    
\begin{eqnarray} 
\label{ra}
\delta^{+}_2D'_{a} =n_2^{(-1)}N'_{a},
\qquad
N'_a=-\frac 1 4 \sigma_2 (\tilde B^{*}_{c\mu\nu}\varepsilon^{\mu\nu\rho\sigma} f^c_{ab}F^{\prime b}_{\rho\sigma}),
\end{eqnarray}
where 
\begin{eqnarray} 
\label{ray}
\tilde B^{*}_{a\mu\nu}=B^{*}_{a\mu\nu}(B^{\prime},A', B^{*\prime},A^{*\prime}),\qquad F^{\prime a}_{\mu\nu}(A')=F^{a}_{\mu\nu}(A), 
\end{eqnarray} 
 $B'=(B^{\prime a}_{\mu\nu}),$ $A^{\prime}=(A^{\prime a}_\mu),$ $B^{*\prime}=(B^{*\prime a}_{\mu\nu}),$ $A^{*\prime}=(A^{*\prime a}_\mu).$ 
The operators $n_2,\sigma_2,\delta_2 $ are given by 
\begin{eqnarray*} 
n_2= B^{\prime a}_{i3}\frac {\delta} {\delta B^{\prime a}_{i3}} +
A^{\prime a}_{\mu}\frac {\delta} {\delta A^{\prime a}_{\mu}}+ 
B^{*\prime a}_{i3}\frac {\delta} {\delta B^{*\prime a}_{ i3}} +
A^{*\prime a}_{\mu}\frac {\delta} {\delta A^{*\prime a}_{\mu}},
\end{eqnarray*}
\begin{eqnarray*} 
\sigma_2= B^{*\prime a}_{i3}\frac {\delta} {\delta B^{\prime a}_{ i3}} +
A^{*\prime a}_{\mu}\frac {\delta} {\delta A^{\prime a}_{\mu}},\qquad
\delta_2= B^{\prime a}_{i3}\frac {\delta} {\delta B^{*\prime a}_{ i3}} +
A^{\prime a}_{\mu}\frac {\delta} {\delta A^{*\prime a}_{\mu}}.
\end{eqnarray*}
One can write 
\begin{eqnarray*} 
N'_a=-\frac 1 8 n_2(\tilde B^{*}_{c\mu\nu}\varepsilon^{\mu\nu\rho\sigma} f^c_{ab}\tilde B^{* b}_{\rho\sigma}) +  W'_a,
\end{eqnarray*}
where
\begin{eqnarray*} 
W'_a=\frac 1 8 n_2(\tilde B^{*}_{c\mu\nu}\varepsilon^{\mu\nu\rho\sigma} f^c_{ab}\tilde B^{* b}_{\rho\sigma}) - \frac 1 4 \sigma_2 (\tilde B^{*}_{c\mu\nu}\varepsilon^{\mu\nu\rho\sigma} f^c_{ab}F^{\prime b}_{\rho\sigma}).
\end{eqnarray*}
Then it follows from (\ref{o10}) and (\ref{ra})
that  
\begin{eqnarray} \label{o100}
M'_{a} =\tilde Y'_{a} -\frac 1 8 \tilde B^{*}_{c\mu\nu}\varepsilon^{\mu\nu\rho\sigma} f^c_{ab}\tilde B^{* b}_{\rho\sigma},
\end{eqnarray} 
where $$\tilde Y'_{a}=Y'_{a}+Z'_{a}, \qquad Z'_{a}= n_2^{(-1)}W'_{a}.$$
Changing variables in \eqref {o100} 
from $(B^{\prime},
A', B^{*\prime},A^{*\prime})$ to
$(B,A, B^{*},A^{*}),$ we get 
\begin{eqnarray} \label{o101}
M_{a} =Y_{a}-\frac 1 8 B^{*}_{c\mu\nu}\varepsilon^{\mu\nu\rho\sigma} f^c_{ab}B^{* b}_{\rho\sigma}, 
\end{eqnarray} 
where 
$$Y_{a}(B, A, B^{*},A^{*})=\tilde Y'_{a}(B^{\prime},
A', B^{*\prime},A^{*\prime}).$$
It remains to check that $\delta Z'_{a}=0.$
By 
the definition of $\delta B^{*a}_{\mu\nu}$ and (\ref{ray}),
$$\delta_2 \tilde B^{*a}_{\mu\nu}=-F^{\prime a}_{\mu\nu},$$ 
and therefore by (\ref {us4}), 
$$Z'_{a}=\delta_2\delta^{+}_2X'_{a}=\delta \delta^{+} X'_{a},$$
where 
\begin{eqnarray*} 
X'_a=\frac 1 8 \tilde B^{*}_{c\mu\nu}\varepsilon^{\mu\nu\rho\sigma} f^c_{ab}\tilde B^{* b}_{\rho\sigma}. 
\end{eqnarray*}

Substituting (\ref{o101}) with $Y_{a}=0$ in (\ref{us22}), we get the extended action
for (\ref{rams})
$$
S=S_0+\int d^4x\left (\frac 1 2 B^*_{a\mu\nu} \varepsilon^{\mu\nu\rho\lambda}
\nabla^a_{\rho b}c^b_\lambda+
c^*_{\mu a}\nabla^{\mu a}_{b}c^{b}+ 
\frac 1 8 B^{*}_{c\mu\nu}\varepsilon^{\mu\nu\rho\sigma} f^c_{ab}B^{* b}_{\rho\sigma}c^{a}\right)
$$ 
in agreement with that of \cite{DGM}.

Let us consider an example of an abelian $L$-th stage reducible theory.
We assume that $R^{a_k}_{ a_{k+1}},$ $k=0,\ldots,L,$ are constants 
and   
\begin{eqnarray}
\label{ry2} 
R^{a_k}_{ a_{k+1}}R^{a_{k+1}}_{a_{k+2}}= 0,\qquad  k=0,\ldots,L-1.
\end{eqnarray}
It is clear that (\ref{us23}) and (\ref{us25}) hold. 
One sees that (\ref{f}) and (\ref{us24}) are satisfied by
\begin{eqnarray*}
M_{a_k}= 0,\qquad k=1,\ldots,L+1.
\end{eqnarray*}
In this case the extended action 
is given by
\begin{eqnarray*}
S=S_0+\sum_{k=1}^{L+1}\phi^*_{a_{k-1}}R^{a_{k-1}}_{a_k}\phi^{a_k}.
\end{eqnarray*}
The off-shell conditions  (\ref{ry2}) hold in abelian $(L+1)$-form gauge theories.
The extended action for these theories is described in \cite{GPS}.

\end{document}